# Thermal contribution to the spin-orbit torque in metallic/ferrimagnetic systems

Thai Ha Pham[1], S.-G. Je[1,2], P. Vallobra[1], T. Fache[1], D. Lacour[1], G. Malinowski[1], M. C. Cyrille[3], G. Gaudin[2], O. Boulle[2], M. Hehn[1], J.-C. Rojas-Sánchez[1*] and S. Mangin[1]

[1] . Institut Jean Lamour, CNRS UMR 7198, Université de Lorraine, F-54011 Nancy, France
[2] . CNRS, SPINTEC, F-38000 Grenoble
[3] . Leti, technology research institute, CEA, F-38000 Grenoble
*juan-carlos.rojas-sanchez@univ-lorraine.fr

## Abstract

We report a systematic study of current-induced perpendicular magnetization switching in W/$Co_xTb_{1-x}$/Al thin films with strong perpendicular magnetic anisotropy. Various $Co_xTb_{1-x}$ ferrimagnetic alloys with different magnetic compensation temperatures are presented. The systems are characterized using MOKE, SQUID and anomalous Hall resistance at different cryostat temperature ranging from 10K to 350 K. The current-switching experiments are performed in the spin–orbit torque geometry where the current pulses are injected in plane and the magnetization reversal is detected by measuring the Hall resistance. The full reversal magnetization has been observed in all samples. Some experimental results could only be explained by the strong sample heating effect during the current pulses injection. We have found that, for a given composition $x$ and switching polarity, the devices always reach the same temperature $T_{switch}(x)$ before switching independently of the cryostat temperature. $T_{switch}$ seems to scale with the Curie temperature of the $Co_xTb_{1-x}$ ferrimagnetic alloys. This explains the evolution of the critical current (and critical current density) as a function of the alloy concentration. Future application could take advantages of this heating effect which allows reducing the in-plane external field. Unexpected double magnetization switching has been observed when the heat generated by the current allows crosses the compensation temperature.

## I.    Introduction

Spin Orbit Torque switching with perpendicularly magnetized material in Hall bar based devices offers a simple and powerful geometry to probe current induced magnetization reversal and had opened a new way to manipulating magnetization at the nanoscale. The underlying physics is quite rich and complex including origin of spin-orbit torque (SOT), interfacial effects and thermal contributions. Magnetization switching by SOT was first observed in heavy metal/ferromagnetic, HM/FM, ultrathin films [1–3]. The torque is mainly related to the spin Hall effect (SHE) [4-10], where the charge current flowing in the heavy metal is converted into a vertical spin current due to the large spin-orbit coupling. This spin current is then transferred to the FM magnetization, which leads to a torque, namely the spin orbit torque. It has been demonstrated for instance using Pt [4–9], Ta [9–12], W [13–16] as HM and FM layers with perpendicular magnetization like CoFeB [7,9–12,14,15], Co [4–6], CoFeAl [16] or (Co/Ni) [8] multilayers. Interface effects can play a key role in the SOT in particular, interfacial spin memory loss [17] and spin transparency [18] which affects the transmitted spin. Furthermore, additional charge to spin current conversion can also occur due to Edelstein effect [19] in Rashba [20] and topological insulator interfaces [19]. There is some attempts to unify a model [21–23] including the aforementioned effects. It might be also an interfacial DMI (Dzyaloshinskii-Moriya interaction) which favors formation of chiral Neel domain wall [24]. It was shown in FM/HM systems that the reversal of the magnetization occurs first by a magnetic domain nucleation followed by a domain wall propagation thanks to SHE and iDMI [8,25–28]. The thermal contribution [6,29] is usually neglected in those experiments.

For possible applications the critical switching current needs to be reduced while maintaining a sufficient thermal stability. In the literature the critical current density to reverse the magnetization, $J_{cc}$, is typically of the order of $\sim 10^{10}$ to $10^{12}$ A/m$^2$ depending on the applied current pulse duration and on the in-plane external magnetic field [4,5,8,30]. $J_{cc}$ is proportional to the magnetization times the thickness of the FM layer ($J_{cc} \propto Mt_F$). Recently, transition metal-rare earth TM-RE ferrimagnetic materials started to attract large attention for spin-orbitronics applications [31–36]. In these ferrimagnetic alloys the net magnetization is given by the sum of the magnetization of the two magnetic sub-lattices (rare earth and transition metal) which are antiferromagnetically coupled. The most advantage of ferrimagnetic materials is that its net magnetization $M$ can be tuned by changing its composition or temperature [37]. As a result, a magnetic compensation point with zero magnetization can occur for a certain alloy concentration, $x_{Mcomp}$, or temperature, $T_{Mcomp}$, where the magnetization of both sub-lattices compensates. Moreover TM-RE thin films are characterized by a large bulk magnetic anisotropy perpendicular to the film plane (PMA) which make easier to integrate TM-RE with different NM materials while keeping large thermal stability [38]. Furthermore, the control of magnetization switching using ultrafast femtosecond laser pulse has been demonstrated recently for various TM-RE materials [39,40]. Those features are encouraging to combine the control of magnetization by both optical and electrical means. Concerning the Spin Orbit Torque (SOT) switching, reports on experiments with TM-RE alloys claim that the spin-orbit torque efficiency reaches a maximum at the magnetic compensation point [31–36], however the critical current is not minimum at this point [33,35,36]. In this study we address the SOT-switching experiments on well characterized //W/Co$_x$Tb$_{1-x}$/Al systems for various concentrations. We demonstrate that thermal effects are keys to explain the current induced magnetization reversal in this system. When the current is injected in the bilayer the Joule heating leads to a large increase of the sample temperature. Using systematic SOT measurements at different temperatures and alloy compositions, we establish that for each concentration *x* the current induced magnetization switching occurs for a unique sample temperature $T_{switch}(x)$. $T_{switch}$ scales with the Curie temperature ($T_C$) of the alloy. Those new findings open new rooms to explore combination of SOT and thermal contribution towards reducing critical current density to reverse *M* and consequently low power consumption applications. In the specific case where $T_{switch}$ is close to $T_{Mcomp}$ an unexpected "double switching" is observed.

## II.   Basic characterization

To study SOT magnetization switching in RE-TM alloys, a model system composed of $Co_xTb_{1-x}$ ferrimagnetic alloys deposited on a tungsten heavy metal with high charge to spin conversion efficiency [13] was considered. The samples were grown by dc magnetron sputtering on thermally oxide Si substrates (Si-SiO$_2$). The full stacks of the samples are Si-SiO$_2$//W(3 nm)/Co$_x$Tb$_{1-x}$(3.5 nm)/Al(3 nm) with $0.71 \leq x \leq 0.86$. The 3 nm thick Al (naturally oxidized and passivated after the deposition) is used to cap the ferrimagnetic layer. The W and CoTb layers have amorphous structure. As described in the introduction, ferrimagnetic alloys like CoTb can show a compensation point at which the Co and Tb moments cancel each other, resulting in zero net magnetization. When the net magnetization of the alloy is parallel (resp. antiparallel) to the magnetization of the Terbium sub-lattice the alloy will be call Terbium rich (resp. Cobalt rich). The samples were characterized by a SQUID-VSM magnetometer and Magneto-optically Kerr effect (MOKE) at room temperature. The SQUID measurements obtained at room temperature are presented in Fig 1a. Magnetization compensation is observed for a concentration $x_{Mcomp}=0.77$ where the coercivity $H_c$ diverges and the net saturation magnetization $M_s$ tends to zero. This value is close to the one reported for bulk and thicker Co$_x$Tb$_{1-x}$ films [37,41] at room temperature. Additionally to the divergence of $H_c$, MOKE measurements show that the Kerr angle rotation changes its sign between Co-rich and Tb-rich samples, which can be explained by the fact that Kerr rotation is mainly sensitive to the Cobalt sub-lattice (see for instance fig. S1 in supplementary material [42]). Both SQUID and MOKE results clearly show that all CoTb films studied have a strong out of plane magnetic anisotropy.

To study Spin Orbit Torque switching, the stacks were patterned by standard UV lithography into micro-sized Hall crosses with a channel of 2, 4, 10 and 20 µm. The results shown are obtained for a width of 20 µm unless otherwise specified. Ti(5)/Au(100) ohmic contacts were defined by evaporation deposition and lift-off method on top of W layers. By measuring the anomalous Hall resistance $R_{AHE}$ of the hall crosses while sweeping the external perpendicular magnetic field $H_z$ at different temperatures, we could determine the magnetic compensation temperature of the samples. Fig. 1b shows the temperature dependence of $H_c$ for 78% of Co. The coercive field $H_c$ diverges around 280 K which determines $T_{Mcomp}$ for this composition.

Moreover, we can observe in the insets that the $R_{AHE}(H_z)$ cycle is reversed for Tb-rich (T<280 K) and Co-rich( T>280 K) phases, namely change of field switching polarity (Field-SP). The latter is due to the fact the Anomalous hall resistance is sensitive to the cobalt sub-lattice. Van der Pauw resistivity measurements leads to a resistivity of W in Si-SiO$_2$//W(3nm)MgO(3nm) of $\rho_W$ = 162 µΩ.cm. Then we could deduce the Co$_{0.72}$Tb$_{0.28}$ resistivity $\rho_{CoTb}$ = 200 µΩ.cm which decreases to 135 µΩ.cm when the Cobalt concentration reaches Co$_{0.86}$Tb$_{0.14}$ in accord with previous results [43]. Despite this trend, the amplitude of $R_{AHE}(H_z)$, $\Delta R_{AHE}$, increases as a function of the Co concentration verifying that $R_{AHE}$ is mainly sensitive to the Cobalt sub-lattice (see also Fig S2 [42]).

### III. Thermally assisted and spin-orbit torque switching

Fig. 2a shows a scheme of a Hall bar along with the conventions used for current injection, voltage probe and directions axes. Typical $R_{AHE}(H_z)$ cycles obtained at room temperature with a low in-plane dc current of 400 µA (charge current density of about 2.4 10$^9$ A/m$^2$ flowing in each layer) for a Tb-rich (resp. Co-rich) sample is shown in Fig. 2b (resp. Fig. 2e). As expected, a change of Field-SP is observed since the alloy net magnetization is parallel to the magnetization of the Cobalt sub-lattice in one case and antiparallel in the other. For the same samples the current-induced switching cycles are shown in Fig. 2c-d (Tb-rich) and 2f-g (Co-rich) with an in-plane bias field of $H_x$=100 mT (Fig. 2c and 2f) and -100 mT (Fig. 2d and 2g). The current injection was performed with pulse duration of 100 µs using a K6221 source coupled to a K2182 Keithley nanovoltmeter. The Hall voltage is measured during the pulse. We have observed the current induced magnetization switching in all the samples for 0.72 ≤ x ≤ 0.86. The Hall resistance amplitudes are the same for the current-switching and the field-switching cycles indicating that the reversal of magnetization is fully achieved in both cases. The data of the full series are shown in Fig. S3 [42]. Sharp current switching are observed and the critical current reduces when $H_x$ increases following similar trends that for ferromagnetic materials [30] as shown in Fig. S4. Remarkably, we observe a full magnetization reversal even for an in-plane field $H_x$ as low as 2 mT. The role of the in-plane field can be understood as the field to balance the iDMI to propagate domain walls which have in-plane magnetization after nucleation of magnetic domains or the field to break the symmetry and to allow for a deterministic switching [8,30]. If the SOT depends on the Co moment, the SOT acts as an effective field $\boldsymbol{H_{SHE}} \propto \boldsymbol{m} \times \boldsymbol{\sigma}$ [24,44] where $\boldsymbol{m}$ is the magnetic moment and $\boldsymbol{\sigma}$ the spin polarization of the spin current $J_s$ injected from the W layer into the CoTb layer. $\boldsymbol{\sigma}$ is along the **y** direction in our measurement geometry (it changes between +y and –y when the direction of the injected current is inverted). $\boldsymbol{m}$ changes its sign upon the change of the in-plane field direction. Then the sign of the Hall cycle vs current, $R_{AHE}(i)$, is reversed when $H_x$ is reversed as observed in Fig 2c-f. Additionally in ferrimagnetic alloys, the effective field $H_{SHE}$ can be reversed if a Co-rich sample is replaced by a Tb-rich one (a schematic is shown in Fig. 2h for $H_x>0$ and $i>0$). The identical effect will be observed if the same sample is kept and the magnetic compensation temperature is crossed. The fact that the samples which have been identified as Co-rich and Tb-rich at room temperature are showing the same current-switching polarity (Fig 2c and 2f) can only be understood if the so call Tb-rich sample has crossed compensation to become Co-Rich. This compensation crossing is due to the Joule heating effect. This assumption was tested by measuring $R_{AHE}(H_z)$ cycles for different applied current pulses on the "Tb-rich" sample shown in Figure 3. We observe that for applied currents $i<19.5$ mA the sign of the cycle demonstrate a "Tb-rich" nature, however for current $i>19.5$ mA the $R_{AHE}(H_z)$ cycles are reversed and demonstrate a "Co-rich" nature. This is clear experimental evidence that for current close to 19.5 mA the device reaches the sample compensation temperature ($T_{Mcomp}$ ~320 K). This demonstrates that the sample is strongly heated during the current-switching experiments. The temperature can be determined by the resistance value as explained in the next section. In Figure 3 the corresponding temperature is shown using color code. We can determine that a temperature of 460 K is reached for 24mA which is the critical current to switch $M$ (Fig. 2c). This current-switching of 24 mA is then obtained for a temperature above $T_{Mcomp}$ which explain why the sign of the $R_{AHE}(i)$ cycle is the one expected for a Co-rich sample. We have performed $R_{AHE}(H_z)$ cycles with intensity current pulse as high as 34 mA (~525 K) where we can observe that the device shows a ferromagnetic hysteresis loop and remain perpendicularly magnetized. $T_C$ is then higher than 525 K.

## IV. Characteristic temperatures of switching

Since we have addressed the reason of the observed switching polarity several questions are arising: i) how much are the devices heated when the switching occurs? ii) Does the temperature at which the switching occurs changes with the initial temperature (temperature at which the experiment is carried out, $T_{cryostat}$) ? iii) How does the switching current and switching temperature depend on composition? And iv) What is the physical meaning of this switching temperature: Angular compensation temperature $T_{Acomp}$? In order to address all those questions we have performed a series of temperature dependence experiments for various samples.

Fig. 4a shows the $R_{AHE}(i_{pulse})$ cycles for $H_x>0$ at different cryostat temperature for W/Co$_{0.73}$Tb$_{0.27}$ (Tb-rich at room temperature). We have observed the Down-Up current-switching polarity at 300 K. For 150 K$\leq T_{cryostat} \leq$250 K a double current switching loop is observed. This type of double current switching can be explained when the switching temperature is close to $T_{Mcomp}$ and its origin will be discussed later in the paper (Fig. 4a shows only the case of 150 K for clarity). For 10 K$\leq T \leq$100 K we observe only Down-Up current-switching polarity (we didn't increase too much the pulse current to avoid burning the device). One can calibrated the real sample temperature at different pulse-current performing the following protocol: i) measuring the resistance of the current channel $R_{channel}(i_{pulse})$ as function of pulse current intensity as shown in Fig. 4b for different cryostat temperatures, and ii) measuring the temperature dependence of the current channel $R_{channel}(T)$ as shown in Fig 4c (for which we use a very low dc bias current of only 400 µA). Interestingly, we observe that for Co-rich current-switching polarity (Down-Up) the device reaches the same resistance (1.373 kΩ) and consequently the same switching temperature $T_{switch}$ = 435 K ± 25 K for this W/Co$_{0.73}$Tb$_{0.27}$. We note that the resistance decreases when $T$ increases which is a feature and confirmation of amorphous materials [45]. We have performed the same protocol for various compositions and different devices. An example for Co-rich sample at room temperature is shown in Fig. 4 d-f (W/Co$_{0.79}$Tb$_{0.21}$) where we also observe that the critical current heats up the device to the same channel resistance (Fig. 4e), so the same $T_{switch}$ (~485 K for this Co$_x$Tb$_{1-x}$ sample) irrespective of the initial temperature. Moreover, on this particular sample $T_{Mcomp}$ is about 500 K and we observe no change of Current-SP even for $T$ as low as 10 K which is well below its $T_{Mcomp}$.

Additionally to the characteristic $T_{switch}$ we have just discussed, one can also investigate the temperature dependence of the critical current as shown in Fig. 5a for a Co$_{0.78}$Tb$_{0.22}$ sample (Co-rich at room temperature). The extrapolation of the linear dependence to zero current is defined as $T^*$. In Fig. 5b is found out that T*~470 K for W/ Co$_{0.78}$Tb$_{0.22}$.

Fig. 6a shows $R_{AHE}(i_{pulse})$ for Co$_{0.72}$Tb$_{0.28}$ performed for a cryostat temperature of 150 K to highlight the observation of the two switching. At lower current (37.5 mA), i.e. lower Joule heating effect, we observed Up-Down current-switching polarity while the second one which occurs at higher current (43 mA) is Down-Up. This can be understood considering that to achieve the first switching the device reaches a temperature below its $T_{Mcomp}$ so the sample is still Tb-rich and the Up-Down current-switching polarity observed is as expected (i.e Fig 2h). If we continue increasing the intensity of the applied in-plane pulse current we overcome $T_{Mcomp}$ and then the perpendicular component of effective torque field $H_{SHE}$ now changes its sign as discussed previously. Consequently, the second observed switching agree well for Co-rich phase (Down-Up). In Fig. 6b is shown the temperature dependence of both switching currents. As discussed, the first (second) switching agrees with a Tb-rich (Co-rich) current-switching polarity and happens for $T<T_{Mcomp}$ ($T>T_{Mcomp}$). The linear extrapolation of both switching currents roughly tends to 350 K (Tb-rich switching) and $T^*$~455 K (Co-rich switching). The value of $T^*$ seems to be slightly higher than $T_{switch}$ (~435 K ± 20 K as determined in Fig 4c).

## V. *T-x* switching phase diagram and conclusions

In figure 7a the different characteristic temperatures of our W/CoTb systems can be plotted in the ($T,x_{Co}$) phase diagram. The determined $T_{Mcomp}$ decreases linearly with the Co concentration as reported for bulk CoTb and thick CoGd films (300 nm) [41,46]. However $T_{switch}$ and $T^*$ increase linearly with the Co-concentration and scale with the Curie temperature $T_c$ thus depending on composition, and independent of initial temperature. It is remarkable that the $T_{switch}$ and $T^*$ are nearly the same, indicating that, to achieve the switching, one has to reach a specific temperature. The three first questions in section IV are answered. Now let's discuss the physical meaning of these switching temperatures. It is clear that for Co-Current SP the temperature of switching is above $T_{Mcomp}$ and below $T_c$. Fig. 7a also shows $T_c$ in bulk CoTb after Hans *et al.* [41]. The angular composition temperature, $T_{Acomp}$, scales with $T_{Mcomp}$. Indeed, typically $T_{Acomp} \sim (T_{Mcomp}+30\ K)$ for $Co_{0.775}Gd_{0.225}$ thick films (300 nm) [46]. This is explained by the relationship between the angular moment $L$, magnetic moment and the gyromagnetic ratio $\gamma$ or Landé g-factor ($L_{TM,\ RE}=M_{TM,\ RE}/\gamma_{TM,\ RE}$ and $\gamma=g\mu_B/h_b$). Therefore in CoTb it is expected that the trends of $T_{Acomp}$ and $T_{Mcomp}$ are similar and they decrease with increasing Co concentration. However $T_{switch}$ increases with Co concentration which indicates that $T_{switch}$ is not scaling with $T_{Acomp}$ or $T_{Mcomp}$. For sake of comparison we plotted the switching-current for experiment performed at room temperature together with the temperature increase $\Delta T = T_{switch}-T_{cryostat}$, due to Joule heating effect, Fig. 7b. We can observe that the temperature is increased between 100 K and 300 K. This variation will increase when we reduce $T_{cryostat}$. Considering that resistivity of both, CoTb and W layers, change similarly with temperature and using the resistivity measured at room temperature we can estimate the critical current density $J_{CC}$ flowing on each layer as displayed in Fig. 7c. We observe that $J_{cc}$ on W is reduced by a factor of ~2 while varying the composition of CoTb. We observe the minimum of $J_{cc}$ at the lower Co concentration measured. This can be explained by the fact that $T_C$ and $T_{switch}$ decrease with decreasing Co-concentration but a relationship between $T_{switch}$ and $T_C$ will need additional studies.

## Conclusions

In conclusion we have fully characterized the current-induced switching experiment in a series of //W(3nm)/$Co_xTb_{1-x}$(3.5nm)/$AlO_x$(3nm) samples. In addition to the SOT effect we demonstrate a strong thermal contribution to achieve the magnetization reversal. For the Co-rich current-switching polarity the device needs to reach the same temperature $T_{switch}$ to achieve the switching. This $T_{switch}$ increases with Co-concentration which then scale with Curie temperature $T_c$. It is then unlikely that $T_{switch}$ corresponds to angular momentum compensation temperature (which scales with $T_{Mcomp}$ decreasing with Co-concentration). Those results highlight the importance of considering thermal contributions in SOT switching experiments and the fact that the spin Hall angle determination might be overestimated when thermal contribution is neglected. The use of resistive W layer increase the heating of the device, reducing strongly the external in-plane needed to assist the SOT. Those results are important for the full understanding of current-induced magnetization switching and may lead the way to new technological applications taking advantages of the rather strong heating

## Acknowledgements

This work was supported partly by the french PIA project "Lorraine Université d'Excellence", reference ANR-15-IDEX-04-LUE. by the ANR-NSF Project, ANR-13-IS04-0008- 01, "COMAG" by the ANR-Labcom Project LSTNM, Experiments were performed using equipment from the TUBE—Daum funded by FEDER (EU), ANR, the Region Lorraine and Grand Nancy. We thank J. Sampaio and S. Petit-Watelot for fruitful discussions.## References

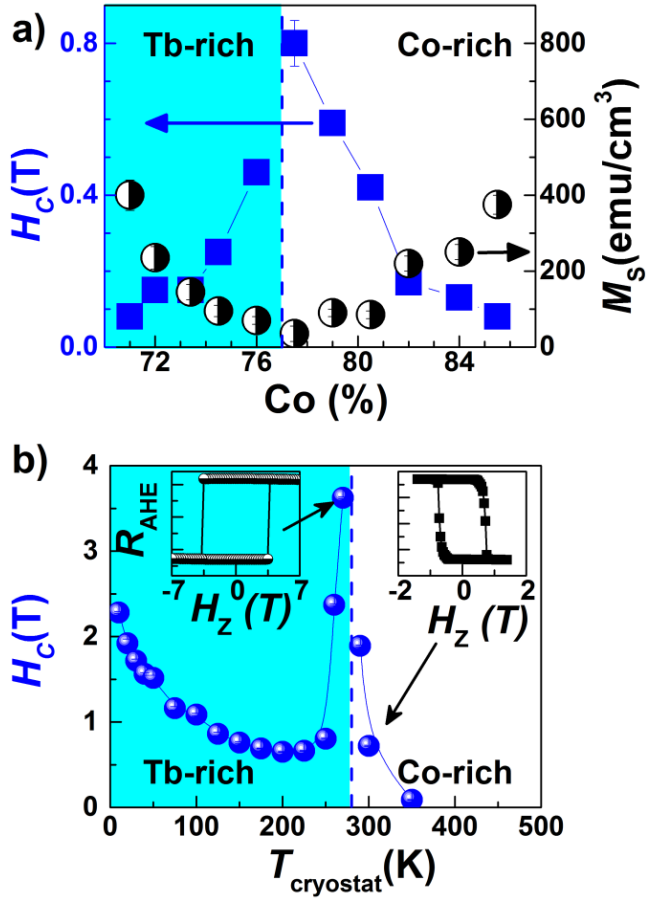

Figure 1. //W(3)/$Co_xTb_{1-x}$ (3.5) /Al(3) : a) Coercive field $H_c$ and saturation magnetization $M_s$ obtained by SQUID measurements *vs.* cobalt concentration at room temperature. Magnetic compensation is observed for $x_{Mcomp}$ ~0.77. b) Temperature dependence of coercivity on Hall bar for x=0.78 showing that $T_{Mcomp}$~280 K for x=0.78. Insets show Hall resistance cycle for a temperature below (left) and above (right) $T_{Mcomp}$. The change of field- switching polarity (Field-SP) evidences that $R_{AHE}$ is mainly sensitive to the magnetization of the Cobalt sub-lattice.

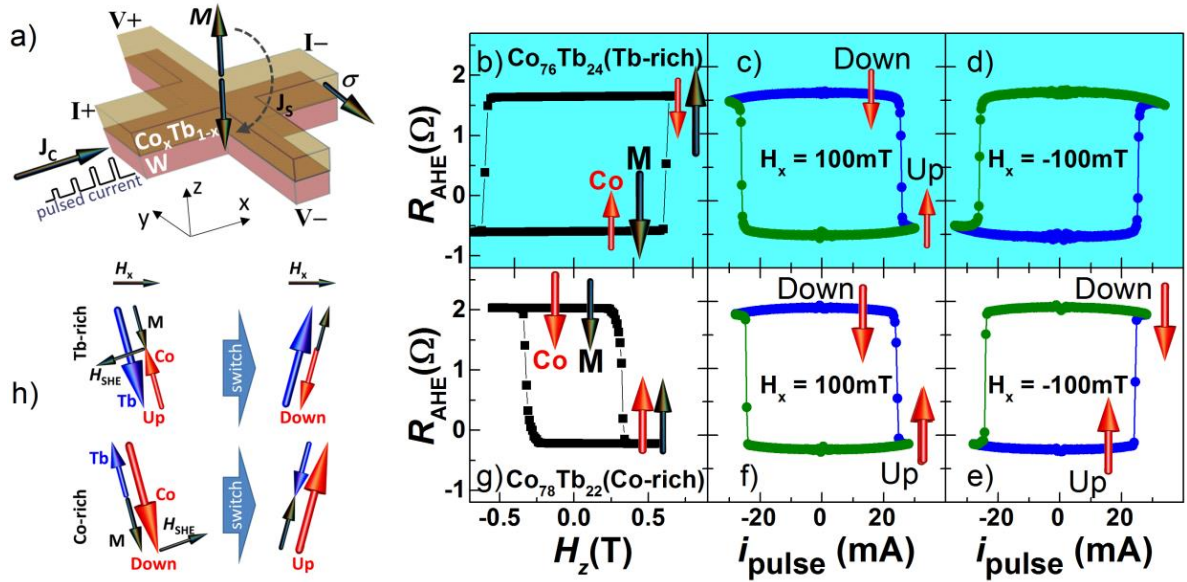

Figure 2. Anomalous Hall effect and Current-induced magnetization reversal at room temperature on Si-SiO$_2$//W(3)/Co$_x$Tb$_{1-x}$ (3.5) /Al(3). a) Scheme of the Hall bar along with geometry used. The spin polarization $\sigma$ is along y axes. (b and g) Sweeping perpendicular field (H||z) with a low dc bias of 400 µA. (c,d,e and f ) Sweeping in-plane current (i$_{pulse}$ ||x) with an in-plane field $H_x$=±100 mT .The width of the channel current is 20 µm. The current switching polarity (current-SP) for $H_x$>0 is Down-Up for Co-rich and Tb-rich samples (c,f). The Current-SP in (c) is opposite than predicted for Tb-rich samples as shown in the schematic in (h). The perpendicular effective torque field $H_{SHE}$ is proportional to $m\times\sigma$ and should lead to a change of the sign in the $R_{AHE}$(i) cycle when changing from Co-rich to Tb-rich phase.

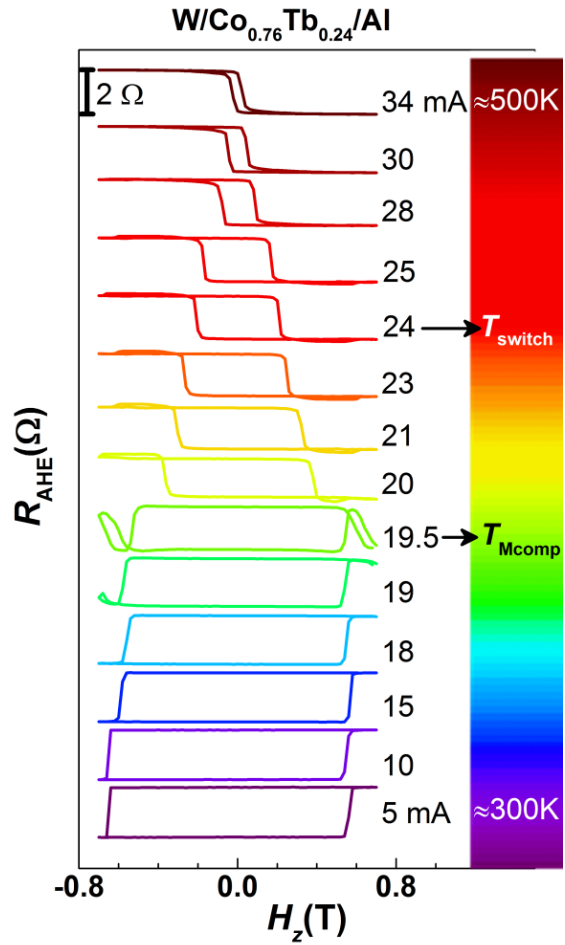

Figure 3. $R_{AHE}(H_z, i_{pulse})$ cycles on Si-SiO$_2$//W(3)/Co$_{0.76}$Tb$_{0.24}$(3.5)/Al(3) Hall bar measured at room temperature. The cycles are vertically offset for clarity. It is observed that for $i <$ 19.5 mA the cycle has a signature corresponding to Tb-rich phase according to our convention. However for $i >$ 19.5 mA the cycle changes their sign and now corresponds to Co-rich phase. It is an evidence of the Joule heating effect when high pulse current is applied. 19.5 mA roughly corresponds to $T_{Mcomp}$. The critical current for this device is about 24 mA. Thus during the electrical switching T$_{device}$>$T_{Mcomp}$>300 K for this Co$_{0.76}$ system.

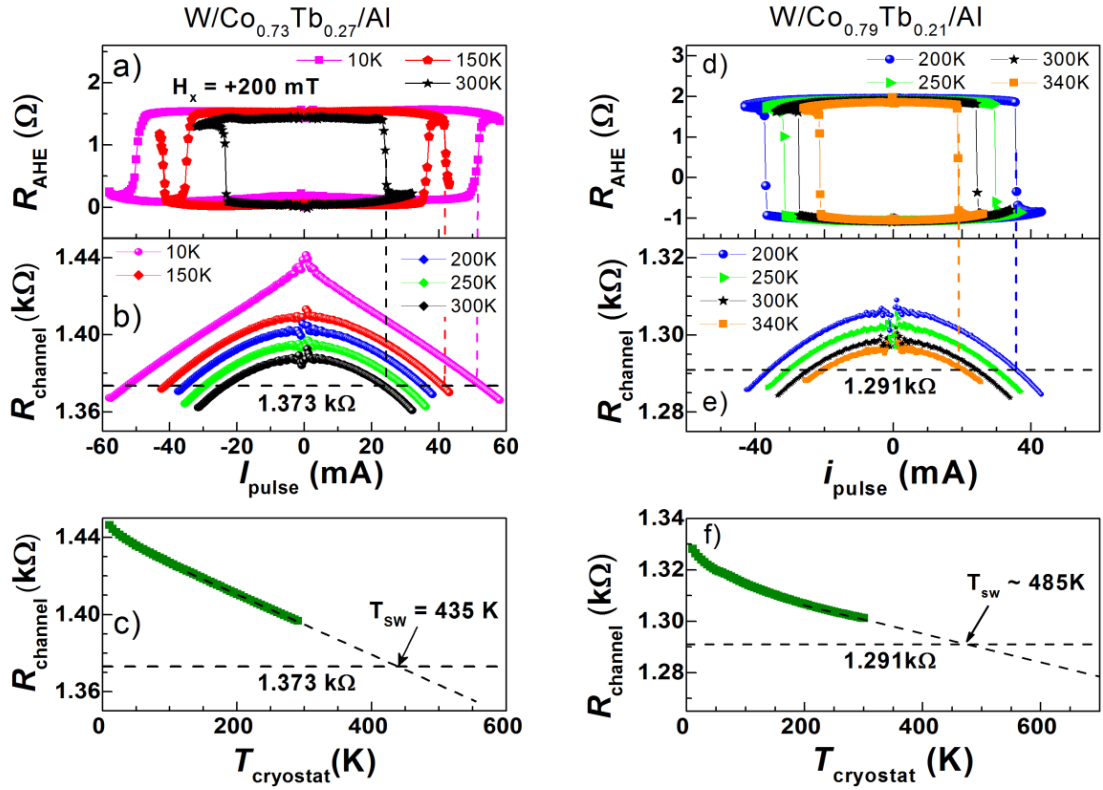

Fig4. a) $R_{AHE}(i_{pulse})$ at different cryostat temperatures. Cycles for x=0.73 (Tb-rich at room temperature). b) $R_{Channel}(i_{pulse})$ and c) $R_{Channel}(T)$ at different cryostat temperatures. The vertical dashed line points out the critical current to reverse $M$. It is observed that independently of the initial temperature, the device always reaches the same value of longitudinal resistance (1373 Ω) which means it reaches the same temperature. The linear extrapolation of $R_{Channel}(T)$ allows us to know the temperature corresponding to the current-induced magnetization reversal . Such a temperature is defined as $T_{switch}$. $R_{Channel}(T)$ is performed with a low bias current of 400 μA. For $Co_{0.73}$ we found that $T_{switch}$= 435 K ± 25 K. d) $R_{AHE}(i_{pulse})$ cycles for x=0.79 (Co-rich at room temperature) at different cryostat temperatures. e) $R_{Channel}(i_{pulse})$ and f) $R_{Channel}(T)$ at different cryostat temperatures. It is also observed that independently of the initial temperature, the device always reaches the same resistance, thus is the same temperature. In this case it corresponds to 1291 Ω and $T_{switch}$ ~ 485 K.

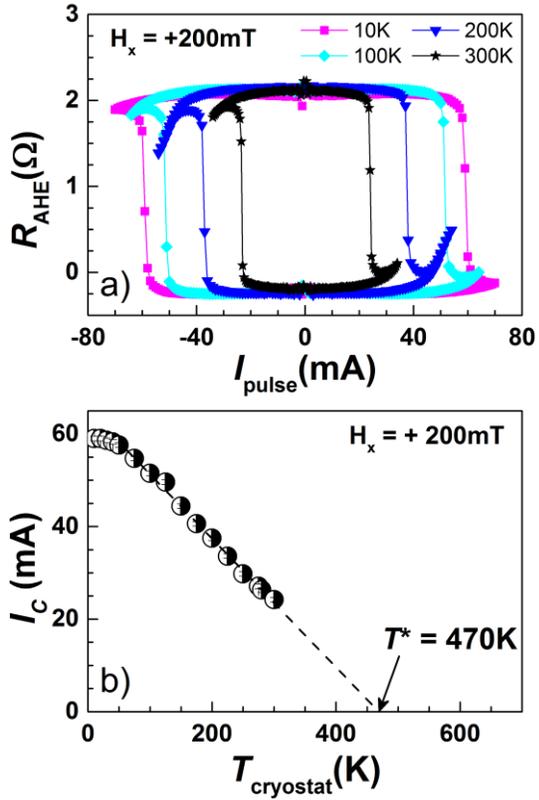

Figure 5. a) $R_{AHE}(i_{pulse})$ at different cryostat temperatures. Cycles for x=0.78 (Co-rich at room temperature). b) The critical current to reverse *M* increases linearly when *T* decrease and saturate for *T*< 50 K. The extrapolation of the linear behavior at higher temperature for zero current is defined as *T*\*.

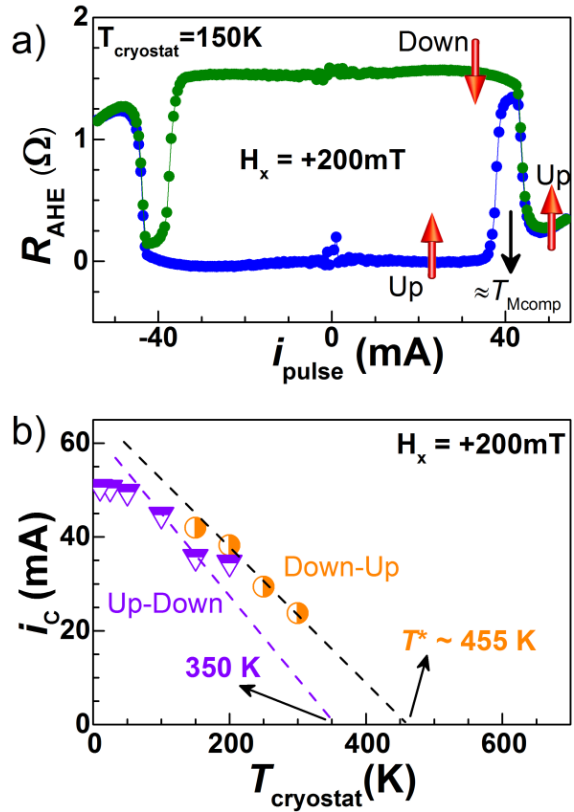

Fig6. a) $R_{AHE}(i_{pulse})$ at 150K for x=0.73 (Tb-rich at room temperature). There are two switchings: i) at lower current it agrees with a Tb-rich switching polarity (Up-Down). ii) The reversal with higher current agrees with a Co-rich switching polarity (Down-Up). b) Temperature dependence of the critical currents for this composition. $T_{Mcomp}$ would be between 350 K and 455 K.

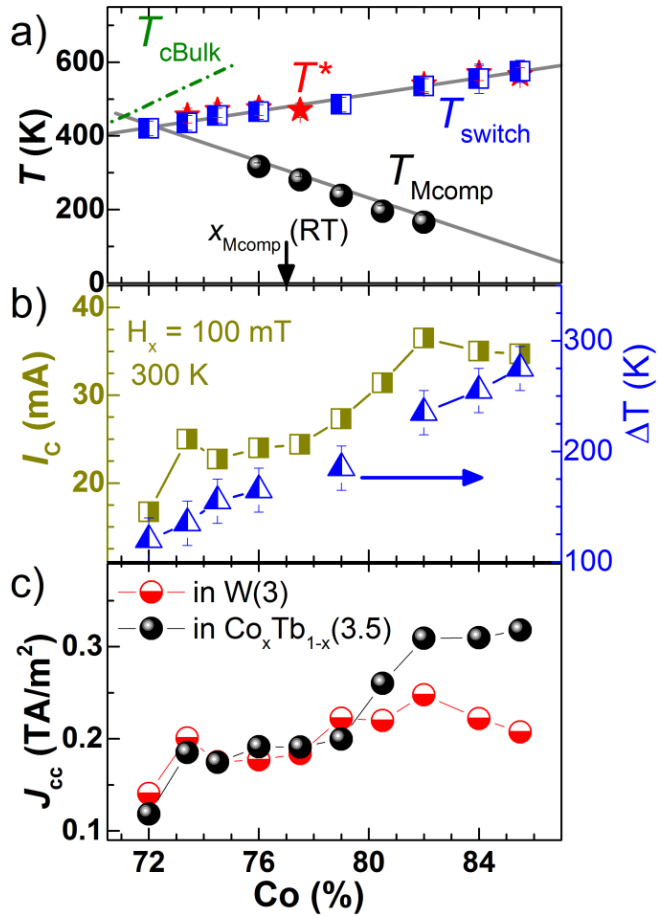

Figure 7. a) Characteristic temperatures as function of Co concentration: $T^*$, $T_{switch}$ and $T_{Mcomp}$ independently measured in Hall bar patterned devices (lines are guides for the eyes). It is observed that $T_{Mcomp}$ follows the same behavior reported for bulk $Co_xTb_{1-x}$ alloys. $T^*$ and $T_{switch}$ have the same trend than that of the Curie temperature Tc. The green dashed line stands for Tc in bulk CoTb after Hans *et al*. [41]. b) The total critical current injected to reverse *M* when the experiment is performed at room temperature (= $T_{cryostat}$), and the variation of temperature $T_{switch}$–$T_{cryostat}$ to reach the switching. c) The critical current density, calculated from b, flowing in W and CoTb layers, respectively.

Supplementary Material

# Thermal contribution to the spin-orbit torque in metallic/ferrimagnetic systems

Thai Ha Pham[1], S.-G. Je[1,2], P. Vallobra[1], T. Fache[1], D. Lacour[1], G. Malinowski[1], M. C. Cyrille[3], G. Gaudin[2], O. Boulle[2], M. Hehn[1], J.-C. Rojas-Sánchez[1*] and S. Mangin[1]

[1] . Institut Jean Lamour, CNRS UMR 7198, Université de Lorraine, F-54011 Nancy, France
[2] . CNRS, SPINTEC, F-38000 Grenoble
[3] . Leti, technology research institute, CEA, F-38000 Grenoble
*juan-carlos.rojas-sanchez@univ-lorraine.fr

## S1- Magneto-optically Kerr effect measurements at room temperature

Figure S1 shows the coercitivity obtained by MOKE of W/$Co_xTb_{1-x}$ thin film as a function of x (the Co-concentration). The results show that coercive field $H_c$ diverges about Co-concentration of 77 % in agreement with SQUID results (Fig. 1a). The insets highlight the opposite sign of Kerr angle rotation when the CoTb alloy change from Tb-rich to Co-rich phase.

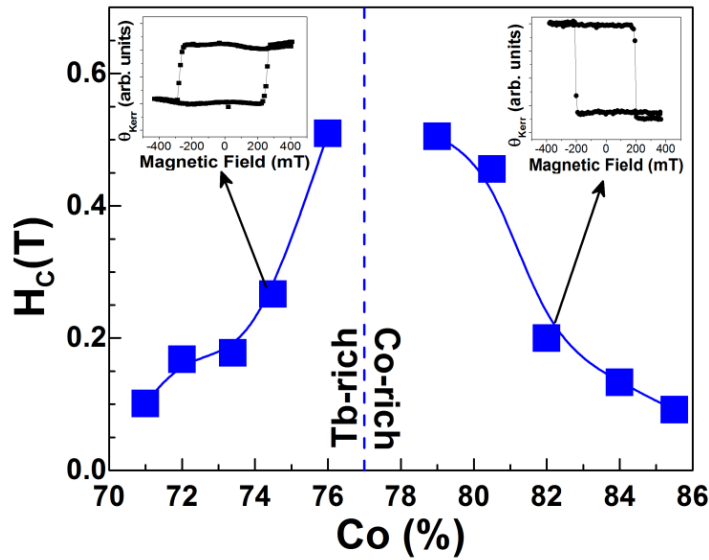

Figure S1. //W(3)/$Co_xTb_{1-x}$ (3.5) /Al(3) : Coercive field $H_c$ obtained by MOKE measurements at room temperature *as a function of the* Cobalt concentration. Insets show raw data of MOKE cycles for a Tb-rich (left) and Co-rich (right) samples.

## S2- Hall resistance amplitude $|\Delta R_{AHE}|$, channel resistance $R_{channel}$, and $Co_xTb_{1-x}$ resistivity $\rho_{CoxTb1-x}$ at room temperature

Figure S2 present the evolution of the Hall resistance amplitude $|\Delta R_{AHE}|$, the channel resistance $R_{channel}$, and the $Co_xTb_{1-x}$ resistivity $\rho_{CoTb}$ at room temperature as a function of the Co-concentration. Despite the decrease of the channel resistance with increasing Co-concentration, we observe that $|\Delta R_{AHE}|$ increases with Co-concentration. Thus corroborate that the Hall resistance is mostly sensitive to Co magnetic moment. The W resistivity, $\rho_W$ = 162 $\mu\Omega$.cm, is of similar order than CoTb alloys.

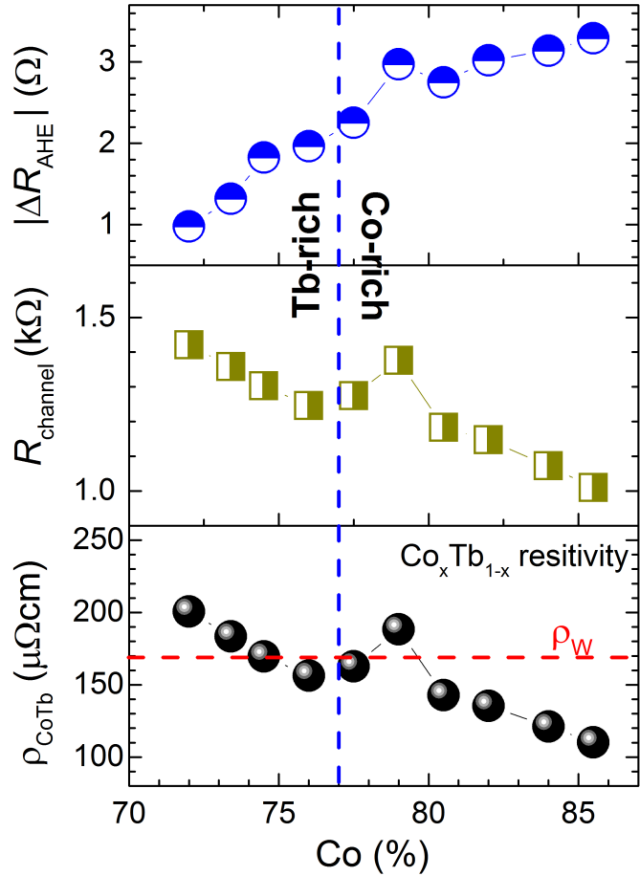

Figure S2. Si-SiO$_2$//W(3)/Co$_x$Tb$_{1-x}$ (3.5) /Al(3): The change of Hall resistance amplitude $|\Delta R_{AHE}|$, channel resistance $R_{channel}$, and Co$_x$Tb$_{1-x}$ resistivity as a function of the Cobalt concentration at room temperature. The vertical blue dashed line points correspond to the magnetic compensation point at room temperature. The horizontal red dashed line shows the value of W resistivity ($\rho_W$).

## S3- Current-switching in the full W/CoxTb1-x series for different in-plane field at room temperature

Figure S3 (S4) presents the current-induced magnetization reversal performed at room temperature on Si-SiO$_2$//W(3 nm)/Co$_x$Tb$_{1-x}$ (3.5 nm) /AlOx(3 nm) with an in-plane field $H_x$=100 mT (5 mT).

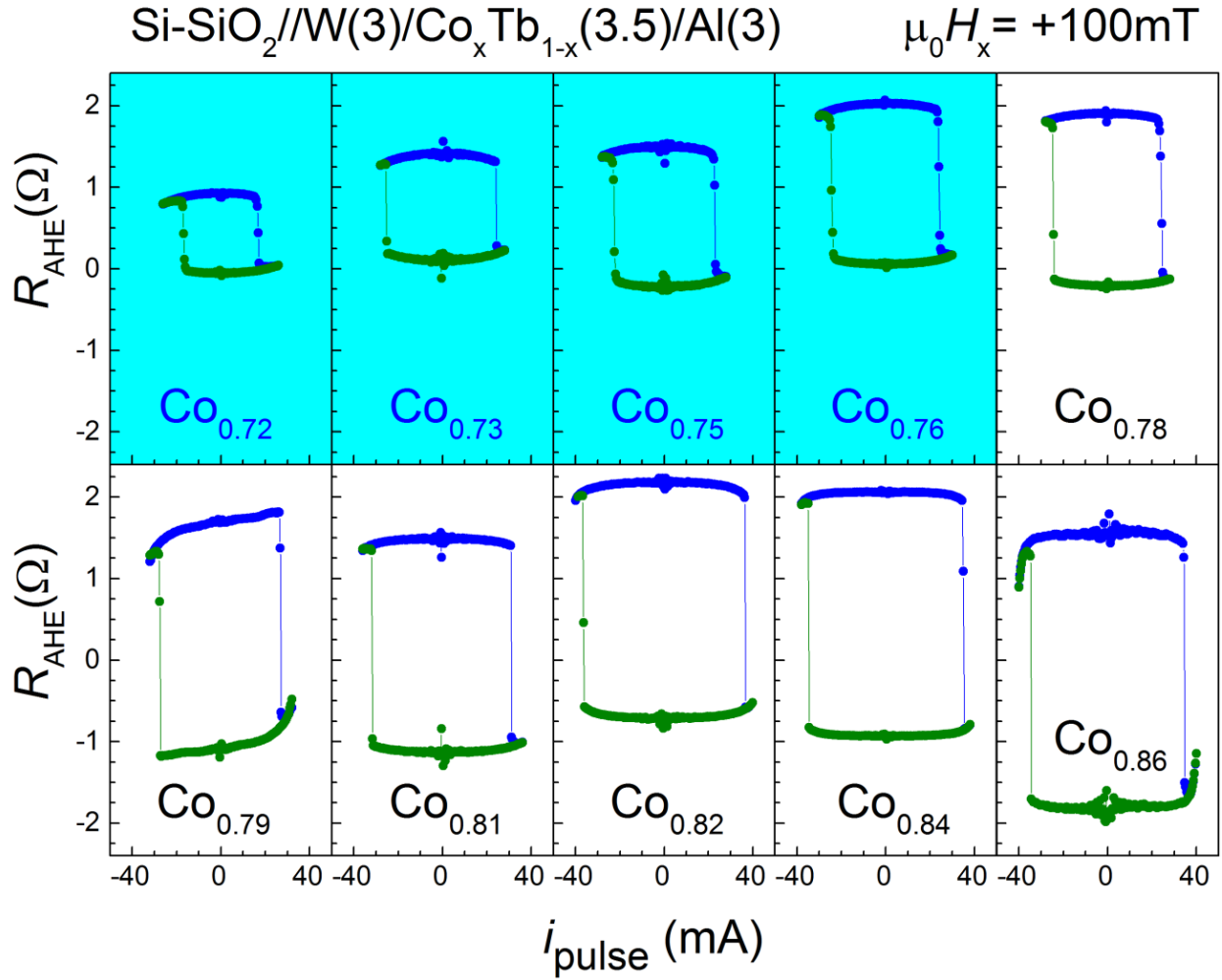

Figure S3. Current-induced magnetization reversal at room temperature: Hall resistance as a function of the injected current measured for various i$_{pulse}$ Si-SiO$_2$//W(3 nm)/Co$_x$Tb$_{1-x}$ (3.5 nm) /AlO$_x$(3 nm), with an in-plane field of $H_x$= 100 mT.

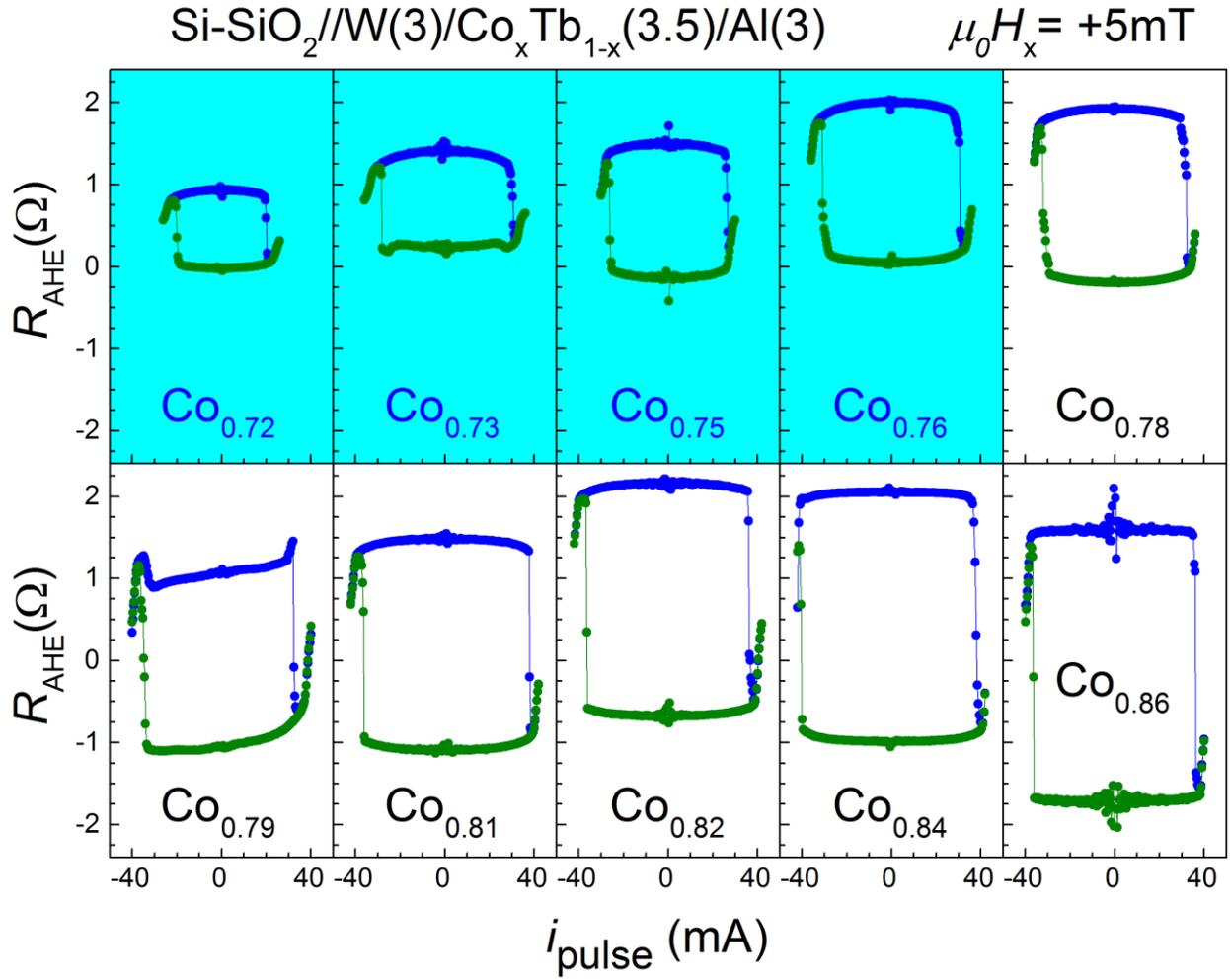

Figure S4. Current-induced magnetization reversal at room temperature: Hall resistance as a function of the injected current measured for various $i_{pulse}$ Si-SiO$_2$//W(3 nm)/Co$_x$Tb$_{1-x}$ (3.5 nm) /AlO$_x$(3 nm), with an in-plane field of $H_x$= 5 mT.

## S4- Power consumption in the W/Co$_x$Tb$_{1-x}$ Hall bar with 20 μm of width at room temperature

Figure S5 shows the electrical power consumption to reverse the magnetization in Hall bar of length L= 100 μm and width of w = 20 μm at room temperature on various Si-SiO$_2$//W(3 nm)/Co$_x$Tb$_{1-x}$ (3.5 nm) /AlOx (3nm) with an in-plane field of 100 mT.

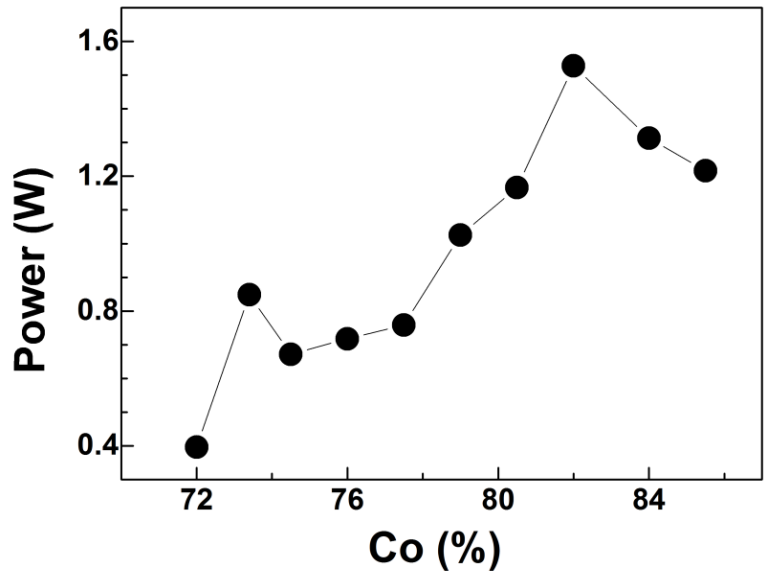

Figure S5. Si-SiO$_2$//W(3)/Co$_x$Tb$_{1-x}$ (3.5) /AlOx(3): The electrical power consumption to switch the magnetization at room temperature of a Si-SiO$_2$//W(3)/Co$_x$Tb$_{1-x}$ (3.5) /AlOx(3) hall bar as a function of Cobalt concentration. The current channel dimensions are length L= 100 μm, and width of w = 20 μm.

## S5- $H_x$-$I$ switching phase diagram in the W/Co$_x$Tb$_{1-x}$

Figure S6 (S7) presents a 2D plot summarizing the current–switching cycles performed under different external in-plane field $H_x$ (phase diagram) at room temperature for a channel width of w= 20 μm (10 μm). Figure S6 are the results obtained for Si-SiO$_2$//W(3 nm)/Co$_{0.86}$Tb$_{14}$ (3.5 nm) /AlOx(3 nm) and fig. S7 for Si-SiO$_2$//W(3 nm)/Co$_{0.84}$Tb$_{0.16}$ (3.5 nm) /AlOx(3 nm).

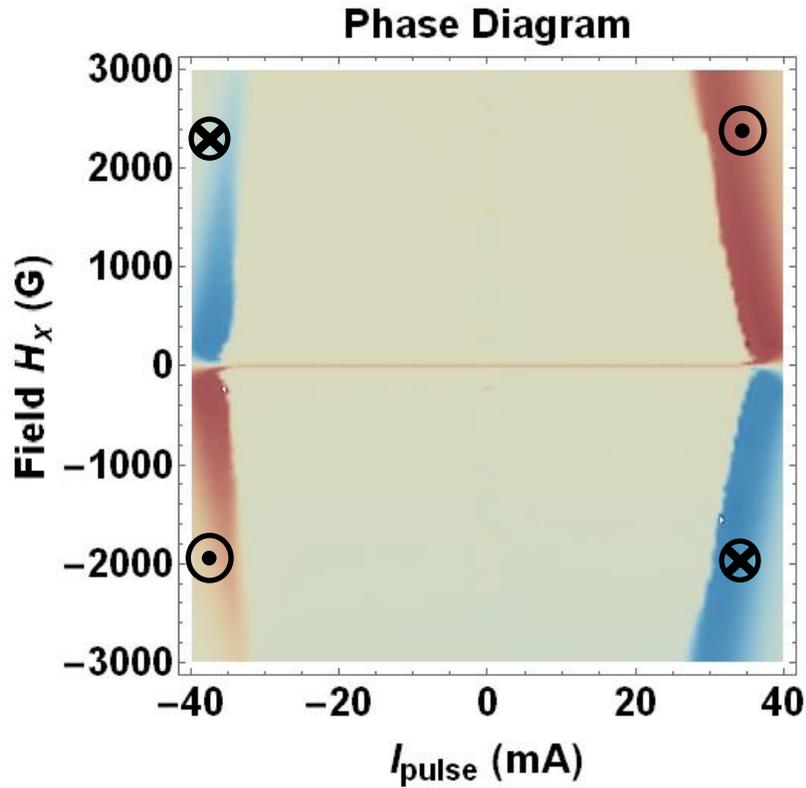

Figure S6. 2D-plot of current–switching cycles performed at room temperature on Si-SiO$_2$//W(3 nm)/Co$_{0.86}$Tb$_{14}$ (3.5 nm) /AlOx(3 nm) for a channel width of w = 20 μm. The $R(i_{pulse}, H_x)$ cycles were carried out with different applied in plane field between −3 kG and +3 kG. The red (blue) color region stand for Up (Down) magnetic configuration according the schematic Hall bar shown in Fig. 2a.

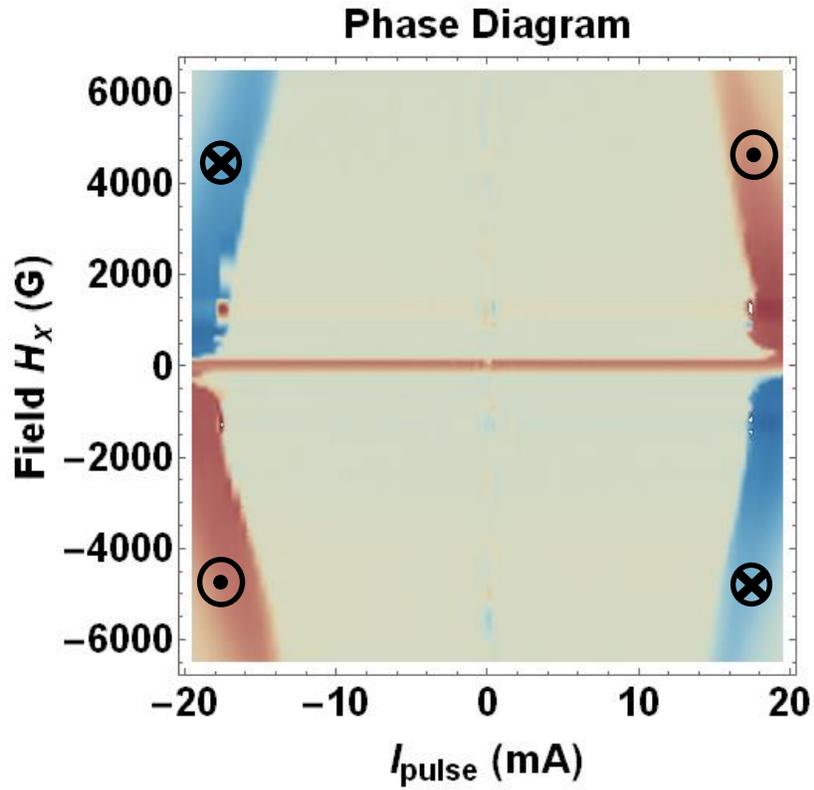

Figure S7. 2D-plot of current–switching cycles performed at room temperature on Si-SiO$_2$//W(3 nm)/Co$_{0.84}$Tb$_{16}$ (3.5 nm) /AlOx(3 nm) for a channel width of w = 10 μm. The $R$(i$_{pulse}$, H$_x$) cycles were carried out with different applied in plane field between –6.5 kG and +6.5 kG. The red (blue) color region stand for Up (Down) magnetic configuration according the schematic Hall bar shown in Fig. 2a.